\documentclass{egpubl}

\JournalSubmission    
 \electronicVersion 

\ifpdf \usepackage[pdftex]{graphicx} \pdfcompresslevel=9
\else \usepackage[dvips]{graphicx} \fi

\PrintedOrElectronic

\usepackage{t1enc,dfadobe}

\usepackage{egweblnk}
\usepackage{cite}
\usepackage[labelfont=bf,textfont=it]{caption}
\usepackage{comment}
\usepackage{amsmath}
\usepackage{amssymb} 
\usepackage{amsfonts}
\usepackage{multirow}
\usepackage{subfigure}
\usepackage{color}
\usepackage{ifthen}
\usepackage{xcolor}
\usepackage[normalem]{ulem} 
\newcommand{\etal}{{\textit{et~al.}}}
\usepackage{enumitem}
\usepackage[encapsulated]{CJK} 
\usepackage{wrapfig}
\usepackage{soul}

\definecolor{gray}{rgb}{0.5,0.5,0.5}
\definecolor{green}{rgb}{0, 0.6, 0}
\definecolor{orange}{rgb}{1, 0.5, 0}
\definecolor{mahogany}{rgb}{0.75, 0.25, 0.0}
\definecolor{purple}{rgb}{0.6, 0, 0.6}
\definecolor{darkgreen}{rgb}{0, 0.3, 0}
\definecolor{orange}{rgb}{1, 0.5, 0.}

\newboolean{revising} 
\setboolean{revising}{true}
\ifthenelse{\boolean{revising}}
{
\newcommand{\ignore}[1]{}
\newcommand{\nothing}[1]{}

\newcommand{\note}[1]{\textcolor{red}{#1}}
\newcommand{\myhl}[1]{#1}
\newcommand{\minorhl}[1]{#1}
\newcommand{\eqhl}[2][yellow]{#2}%

\newcommand{\iccmt}[1]{\textcolor{purple}{ichao: #1}}

\newcommand{\chinky}[1]{\textcolor{black}{#1}}
\newcommand{\chireplace}[2]{\textcolor{black}{#2}}
\newcommand{\robin}[1]{{#1}}
}
{
\newcommand{\note}[1]{}
\newcommand{\iccmt}[1]{}

\newcommand{\chinky}[1]{{#1}}
\newcommand{\chireplace}[2]{{#2}}
\newcommand{\robin}[1]{{#1}}

\newcommand{\ignore}[1]{}
\newcommand{\nothing}[1]{}
}

\usepackage{algorithm}
\usepackage{algorithmicx}
\usepackage{algpseudocode}
\algnewcommand\algorithmicinput{\textbf{Input:}}
\algnewcommand\INPUT{\item[\algorithmicinput]}
\algnewcommand\algorithmicoutput{\textbf{Output:}}
\algnewcommand\OUTPUT{\item[\algorithmicoutput]}
\algnewcommand\algorithmicforeach{\textbf{for each}}

\algdef{S}[FOR]{ForEach}[1]{\algorithmicforeach\ #1\ \algorithmicdo}
\algrenewcommand{\alglinenumber}[1]{\color{red!80!blue}\footnotesize#1:}

\algnewcommand\Func[2]{\textcolor{green}{#1}\textcolor{green}{(#2)}}
\algnewcommand\Insert[2]{Insert {#1} to #2.}
\algnewcommand\Input[1]{\State \textbf{Input: } #1}
\algnewcommand\Output[1]{\State \textbf{Output: } #1}

\newcommand{\ie}{i.e.}

\newcommand{\figname}{Figure}
\newcommand{\tabname}{Table}
\newcommand{\secname}{Section}

\newcommand{\eqname}{Eq.}

\title[ZomeFab: Cost-effective Hybrid Fabrication with 
Zometools]
      {ZomeFab: Cost-effective Hybrid  Fabrication with 
      Zometools} 

\makeatletter
\renewcommand*{\@fnsymbol}[1]{\ifcase#1\or*\else\@arabic{\numexpr#1-1\relax}\fi}
\makeatother

\author[I-Chao Shen \& Ming-Shiuan Chen \& Chun-Kai Huang \& Bing-Yu Chen]
{\parbox{\textwidth}{\centering I-Chao Shen~
        Ming-Shiuan Chen ~ Chun-Kai Hunag ~ Bing-Yu Chen
        }
        \\
{\parbox{\textwidth}{\centering National Taiwan University
       }
}
}

\begin{document}

\maketitle
\begin{abstract}
In recent years, personalized fabrication has received considerable attention because of the widespread use of consumer-level three-dimensional (3D) printers. However, such 3D printers have drawbacks, such as long production time and limited output size, which hinder large-scale rapid-prototyping.
In this paper, for the time- and cost-effective fabrication of large-scale objects, we propose a hybrid 3D fabrication method that combines 3D printing and the \textit{Zometool} construction set, which is a compact, sturdy, and reusable structure for infill fabrication. The proposed method significantly reduces fabrication cost and time by printing only thin 3D outer shells. In addition, we design an optimization framework to generate both a Zometool structure and printed surface partitions by optimizing several criteria, including printability, material cost, and Zometool structure complexity. Moreover, we demonstrate the effectiveness of the proposed method by fabricating various large-scale 3D models.


\end{abstract}
\section{Introduction}
\label{sec:introduction}
The recent widespread adoption of consumer-level three-dimensional (3D) printers spurred the growth of academic and industrial fabrication applications.
However, several drawbacks, including long production time, limited output size, and 
high material costs, persist. Many 
approaches have been
proposed to address these shortcomings.
To reduce production time and material costs, 
modern 3D printers allow users to vary the fill rate.
Furthermore, various internal patterns have been proposed~\cite{Lu:2014:BSW} to reduce the amount of material used without sacrificing structural soundness. 
In addition, some methods \cite{Luo:2012:CPM, Hu:2014:APS, Vanek:2014:PMVO} 
have been proposed~for large-scale 
fabrication, most of which are based on the principle of partitioning large objects into smaller components that  are compatible with the printer's output size.

Large-scale fabrication using extant methods remains expensive.
\ignore{ are still not sufficient for large-scale fabrication.}
To substantially reduce fabrication costs, we propose a hybrid fabrication method that integrates the use of 3D printing materials for the external shape with the use of a supporting structure for the internal volume. 
The inner structure must satisfy several criteria; for example, the structure must be   (1) easy to assemble, (2) reusable, and (3) robust. These criteria ensure that the resulting fabrication 
is simple to build while also being structurally reliable. 
\emph{\textit{\textit{\textit{}}} Zometool}~\cite{davis2007mathematics} satisfies these requirements 
for coarse fabrication.
 \textit{\emph{Moreover, Zometool} }has other favorable characteristics, such as stability and lightness, which facilitate large-scale fabrication, and structural modularity, which facilitates fast fabrication.
In this work, we develop a computational method that integrates 3D printing with the Zometool structure to reduce the time and material costs of large-scale fabrication.

For a given 
3D shape, we design an optimization process to synthesize the inner Zometool structure such that  shape similarity and structural complexity are optimally traded off.
Through simulated annealing, we effectively explore the large structure space.
Next, using the optimized Zometool structure, we hollow out the shape to obtain the outer shell and partition such that several criteria, including simplicity and printability, are satisfied.
These criteria are formulated as a single \myhl{multiclass labeling}
problem and solved using a graph-cut algorithm~\cite{boykov:2004:experimental}.
We then design a particular type of connector and optimize its positions for assembling the inner Zometool structure and the outer shell.

This paper makes two primary contributions:
\begin{itemize}
\item We propose an optimization framework to synthesize the inner Zometool structure that replaces solid printed materials in large-scale fabrication.
\item We design and print a special connector and optimize its layout for effectively combining the inner Zometool structure and the outer printed shell.
\end{itemize}

\section{Related Work}
\label{sec:relatedwork}
\subsection{Computational Fabrication}
In recent years, computational fabrication has attracted considerable attention in the fields of computer graphics and human\chinky{-}computer interaction \cite{Shamir:2016:CTP}.
Numerous approaches have been proposed for fabricating shapes 
that satisfy various objectives (e.g., maintaining balance~\cite{Prevost:MIS:2013,SpinIt:Baecher:2014}, reducing size~\cite{Luo:2012:CPM}, strengthening \chireplace{structure}{structural} soundness~\cite{Zhou:2013:WSA}, and generating specific sounds~\cite{Umetani:2016:PIR}) while using various materials and building blocks (e.g., Lego~\cite{Luo:2015:LOL}, planar slices~\cite{Cignoni:2014:FMJ},  and interlocking puzzles~\cite{Song-2012-InterCubes, alemanno2014interlocking}).
A detailed review of these approaches is available in \cite{BCMP18}.

Despite the development of assisting tools and algorithms, 3D printers still have such drawbacks as long production time, excessive material use, and limited output size.
To reduce the consumption of print materials, Huang~\etal~\cite{Huang:2016:FRF} and Wu~\etal~\cite{Wu:2016:PAM} have designed devices and algorithms that print shapes in wireframe form.
In addition, several studies have developed different types of internal structures, such as~the skin-frame structure~\cite{Wang:2013:CPO} and the honeycomb-like structure~\cite{Lu:2014:BSW}.
For the 3D printing of large structures, Luo~\etal~\cite{Luo:2012:CPM} developed an iterative planar-cut method with the aim of fitting decomposed parts within the 3D printing volume while considering factors such as assemblability and aesthetics.
Yao~\etal~\cite{Yao:2015:LPP} proposed a level-set framework for 3D shape partition and packing.
Compared with the aforementioned works, our method fabricates a shape using both Zometool structures and 3D-printed pieces, thus reducing the fabrication time and cost given the reusability of Zometool structures. 

Several studies have focused on hybrid fabrication, for example, CofiFab~\cite{Song-2016-CofiFab}, \myhl{Universal Building Block~\mbox{\cite{Chen:2018:FUB}}}, and faBrickation~\cite{Mueller:2014:WPP}.
Our work primarily differs from \cite{Song-2016-CofiFab} and \cite{Chen:2018:FUB} in the use of Zometool for fabricating the inter structure. 
The advantages of Zometool are as follows: (i) Zometool is easy to obtain and manipulate compared with the customized laser-cut shapes used in CofiFab; (ii) Zometool elements are reusable, making them cheaper than the laser-cut shapes; and (iii)\ Zometool structures are relatively sparse, thus necessitating relatively few building elements and enabling material savings. 
Different inter structures necessitate the use of different processes and algorithms. For example, the applications of faBrickation differ from those of our method. 
In faBrickation, Lego bricks are used as the main building blocks to realize shapes, whereas 3D-printed parts are used to fill the smaller components that are difficult to build using Lego bricks.

\subsection{Zometool Design and Modeling}
Zometool is a mathematically precise plastic construction set used for building a myriad of geometric structures~\cite{davis2007mathematics}; it can be used to visualize simple polygons as well as model complex structures such as DNA molecules.
Zometool dates back to the 1960s, when it was first used as a simple construction system inspired by Buckminster Fulleresque geodesic domes. Its application has since evolved to satisfy the requirements of versatile modeling.
Although Zometool can be used to construct complex structures, it is not an intuitive tool for new users, and its use can also be time consuming.
Various tools have been developed to help users design Zometool structures, such as vZome\footnote{http://vzome.com} and ZomeCAD\footnote{http://www.softpedia.com/get/Science-CAD/ZomeCAD.shtml};
these systems provide multiple methods to grow structures.
However, building a complex shape remains difficult because these systems do not provide suggestions to the users, such as what item to use next.

To address these concerns, several studies have proposed automatic construction through the use of computational methods.
Zimmer~\etal~approximated and realized freeform surfaces automatically by using Zometool.
Zimmer and Kobbelt~\cite{zimmer:2014:tvcg} adopted a growth strategy that entails the use of incremental panels to approximate the desired surface without self-collisions.

\section{Overview}
\label{sec:overview}
\begin{figure*}[ht]
\centering
\includegraphics[width=0.9\linewidth]{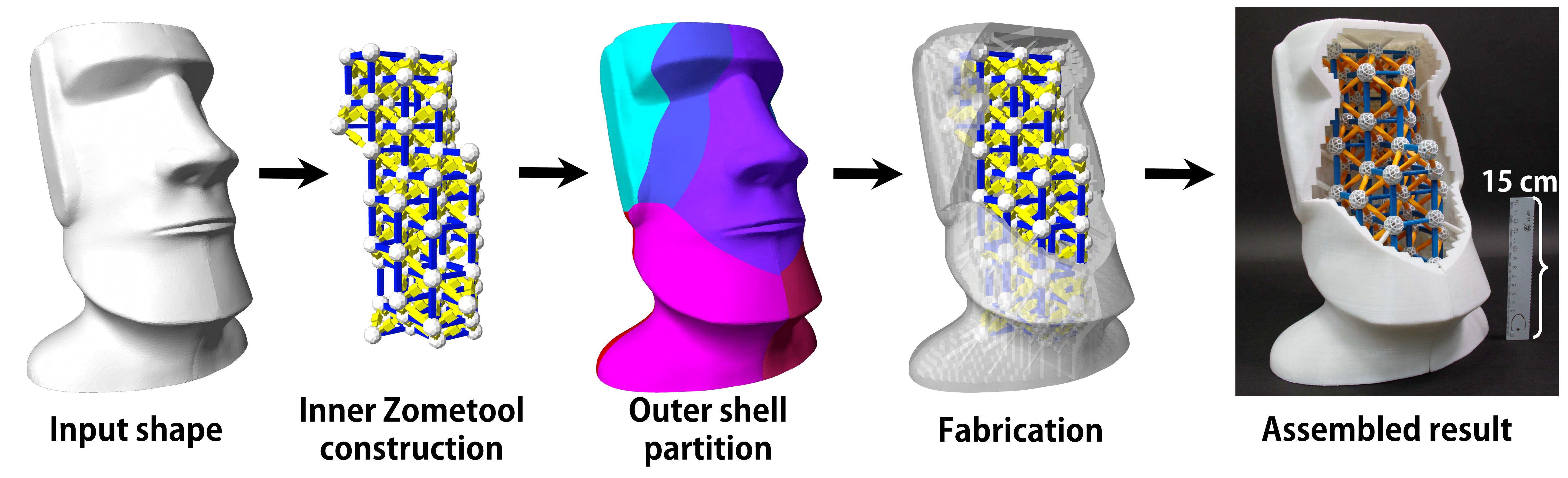} 
\caption{
For a given input shape, we first optimize the inner Zometool structure (\secname~\ref{sec:Zometool}). 
Guided by the 
Zometool structure, we then partition the outer shell (\secname~\ref{sec:surf_part}) and generate connectors for assembling them (\secname~\ref{sec:fab}). 
The final fabricated result is obtained by assembling both assembled Zometool structure and printed outer shell. 
}
\label{fig:result-pipeline}
\end{figure*}

\minorhl{Given a 3D shape \mbox{$M$}}, the proposed method automatically generate\chinky{s} an inner Zometool structure and outer 3D-printed shells.
This method has the following features:
\begin{enumerate}
\item \textbf{Large Objects.} The proposed method is designed for fabricating large-scale objects whose volume exceeds the printing volume of most current consumer-level 3D printers.
\item \textbf{Fabricability.} Each segment of the outer shell can be printed such that it fits inside the volume of consumer-level 3{D} printer\chinky{s}.
\item \textbf{Assemblability.} The inner Zometool structure can be easily assembled and connected to the outer printed shells by using
\chireplace{special}{specifically} designed connectors.
\item \textbf{Cost-effectiveness.} Our method maximizes the volume of the inner structure and minimizes the amount of printing materials used. 
\end{enumerate}

\figname~\ref{fig:result-pipeline} illustrates our method. For a given input shape, we voxelize the inner volume of the input mesh to realize an initial Zometool structure.
The objective is to grow the Zometool structure such that it maximizes the inner volume, thus reducing the amount of printing materials necessary.
To this end, we design an optimization framework through simulated annealing and design several local operations to explore the optimized Zometool structure space (see \secname~\ref{sec:Zometool}).

Next, to print the outer shell with the appropriate shape, we  (i) partition the shape into pieces such that each piece  fits inside the printing volume, (ii) place connectors at feasible locations so that the inner Zometool structure and the outer shell can be connected robustly, and (iii) maintain the salient region\chinky{s} intact.
We formulate this partition problem as a multi-label problem and solve it by using a graph-cut algorithm
under the  following considerations: each triangle must be connected to its closest Zomeball, but the number of partitions must also be reduced without sacrificing the integrity of the salient regions.
To further regularize the resulting partitions, we apply a support vector machine (SVM) algorithm to  determine the hyperplane  between different labels. We then use this hyperplane as the cut plane to separate the mesh (see \secname~\ref{sec:surf_part}).

Before separating the mesh using the aforementioned cut plane, we generate the inner surface by voxelizing the inner volume and combine it with the outer mesh as the solid mesh. 
Subsequently, we apply all the cut planes to the solid \ignore{new }mesh and obtain the distinct pieces. 
Finally, we design a special connector and connect the printed pieces to the inner Zometool structure to form the designed large-scale object (\secname~\ref{sec:fab}).
\section{Zometool construction}
\label{sec:Zometool}




\begin{wrapfigure}{R}{0.3\textwidth}
\centering
\includegraphics[width=0.29\textwidth]{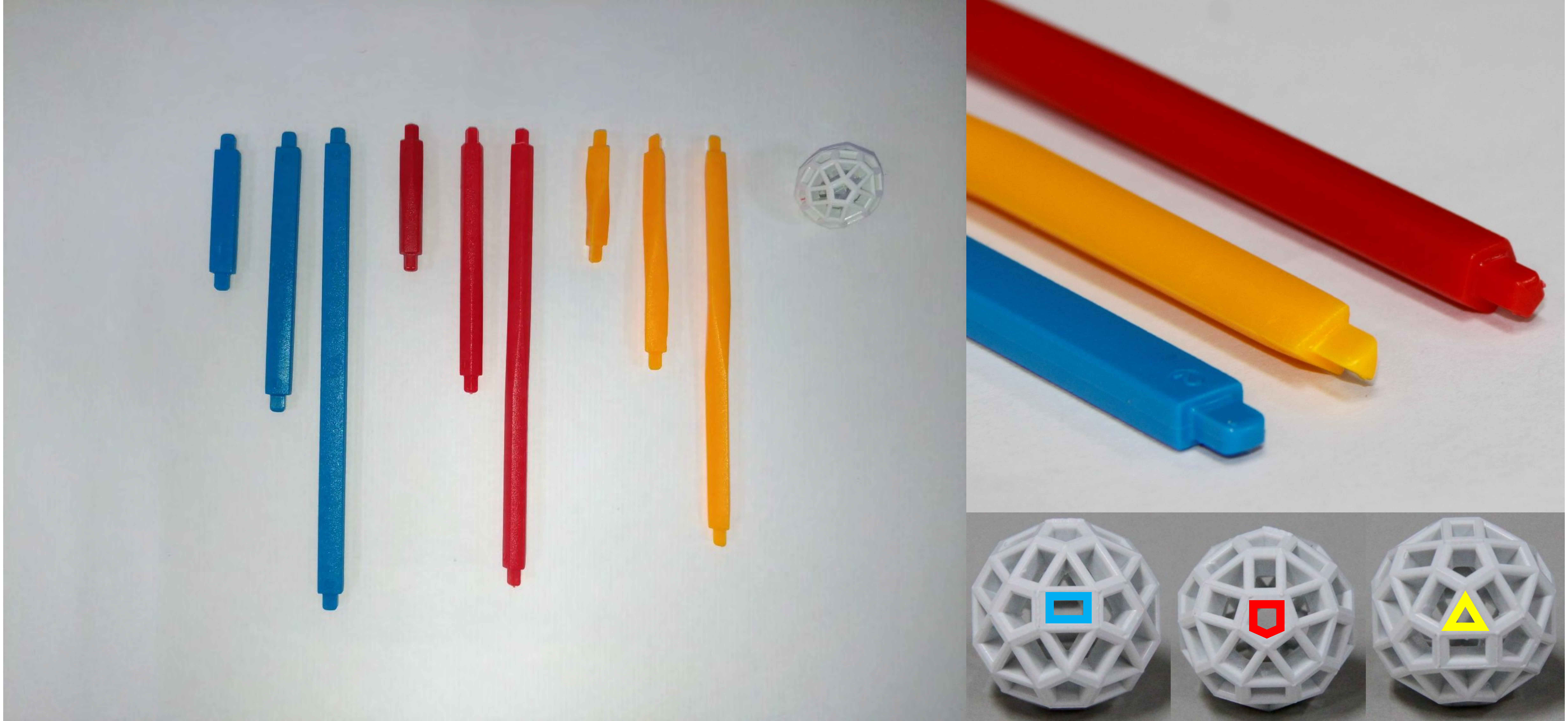}
\label{fig:Zometool}
\end{wrapfigure}

\subsection{Introduction to Zometool}
Zometool is widely used as an educational toy that replicates complex scientific structures, such as chemical structures.
The rods in the standard Zometool system have three types of struts: blue for rectangle, red for pentagon, and yellow for triangle.
Each strut comes in three lengths, with length here defined as the distance from the center of a ball on one end to the center of a ball on the other end (see inset).
We denote $(b_0$, $b_1$, $b_2)$ as the three lengths of the blue struts; similarly, $(r_0$, $r_1$, $r_2)$ and $(y_0$, $y_1$, $y_2)$ represent the red and yellow struts, respectively.
\myhl{} 
The ratio of the lengths follows the golden ratio, $\gamma = \frac{1+\sqrt{5}}{2}$. 
For example, for the blue struts, $b_1 = b_0 \cdot \gamma$ and $b_2 = b_0 + b_1$. 
Moreover, the relative length ratio of the yellow and blue struts and that of the red and blue struts differ: $y_i = \frac{\sqrt{3}}{2} \cdot b_i$ and $r_i = \frac{\sqrt{2 + \gamma}}{2} \cdot b_i$.
Each Zomeball has\chireplace{totally}{} 62 slots, namely 30 rectangular, 12 pentagonal,  and 20 triangular slots. 
Please refer to~\cite{davis2007mathematics} for more details on the mathematical model of Zometool.


\begin{wrapfigure}{R}{0.1\textwidth}
\centering
\includegraphics[width=0.1\textwidth]{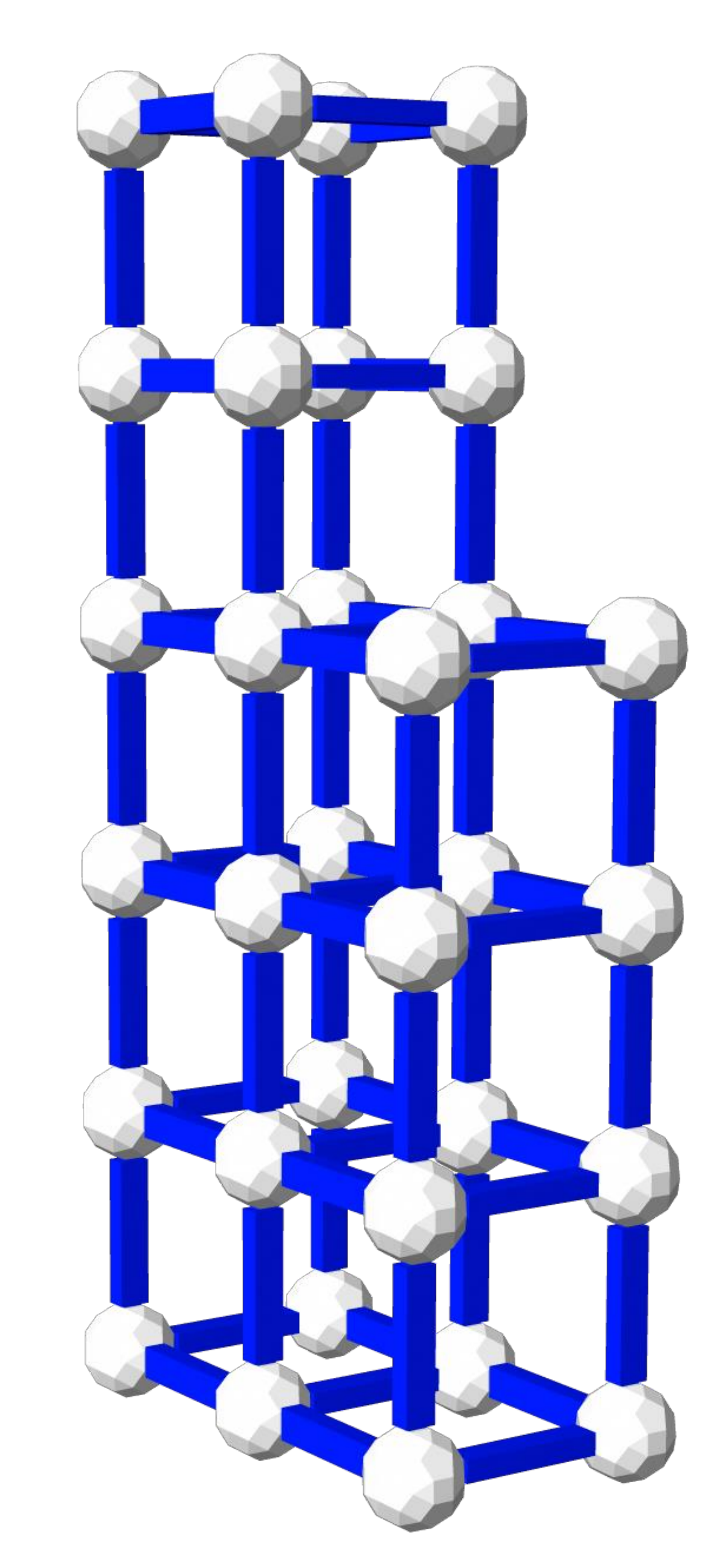}
\label{fig:Zometool_cont}
\end{wrapfigure}

\subsection{Initialization}
Many complicated structures can be assembled using Zometool. Nevertheless, the assembly complexity and time increase rapidly with the number of Zometool items (i.e., struts\chinky{} and ball\chinky{s}) used.
A simple and repeating unit structure is used to fabricate the initial structure, following which the rest of the structure is built outward from the initial structure to achieve a good fit with the outer surface.
After experimenting with different basic structures (e.g., cube\chinky{s}, triangular pyramid\chinky{s}, square\chireplace{-based}{} pyramid\chinky{s,} and pentagonal pyramid\chinky{s)}, we select the cube as the unit building block for this study because it has the shortest assembly time.
The major difference between our study and that of Zimmer~\etal~\cite{zimmer:2014:Zometool} is that we choose $b_0$ (the length of the shortest blue rod) as the edge length of the cube; this is because the smaller is the cube, the higher is the fitting rate, and the closer are the inner and outer surfaces (inset shows sample initialization).

\subsection{Problem Formulation}
We measure the quality of the Zometool structure $\mathbf{Z}$ \ignore{model }with energy $E$ in terms of four 
quality measurements:
\begin{align}
\label{eq:sa_energy}
E(\mathbf{Z})&=w_{\text{fid}}\cdot E_{\text{fid}}(\mathbf{Z}) + w_{\text{reg}} \cdot E_{\text{reg}}(\mathbf{Z}) \nonumber \\ 
&+ w_{\text{val}}\cdot E_{\text{val}}(\mathbf{Z}) + 
w_{\text{sim}} \cdot E_{\text{sim}}(\mathbf{Z}),
\end{align}
We set $w_{{fid}}$ = 1.0, $w_{{reg}}$ = 100.0, $w_{{val}}$ = 1.0, $w_{{sim}}$ = 1.0 for all examples in this paper.

\subsubsection{Shape fidelity}
\label{sec:opt_distance}
To better represent the input shape $S,$ the outermost nodes must stay close to the surface of $S$; that is, the distance from the outermost nodes to the surface of $S$ should be minimized.
Thus, the distance from $\mathbf{Z}$ to $S$ is integrated over the outermost nodes:
\begin{align}
E_{\text{fid}}(\mathbf{Z}) = \frac{1}{|N_{out}|\cdot d^2_{\text{norm}}} \sum_{i=1}^{|N_{out}|} \|p_i - \delta(p_i) \|^2 \cdot (1+F(p_i)),
\end{align}
\myhl{where \mbox{$N_{out}$} is the set of outermost nodes and \mbox{$\delta(p_i)$} is the point on \mbox{$S$}} nearest to the outermost nodes. The normalization factor $d_{\text{norm}}$ is \myhl{set as \mbox{$b_0$} to retain the energy within \mbox{$[0,1]$}}.
We follow \cite{zimmer:2014:Zometool} and define the term $F(p)$ as the forbidden zone that penalizes node points lying too far away from the surface. 

\begin{figure}[ht]
\centering
\includegraphics[width=0.8\linewidth]{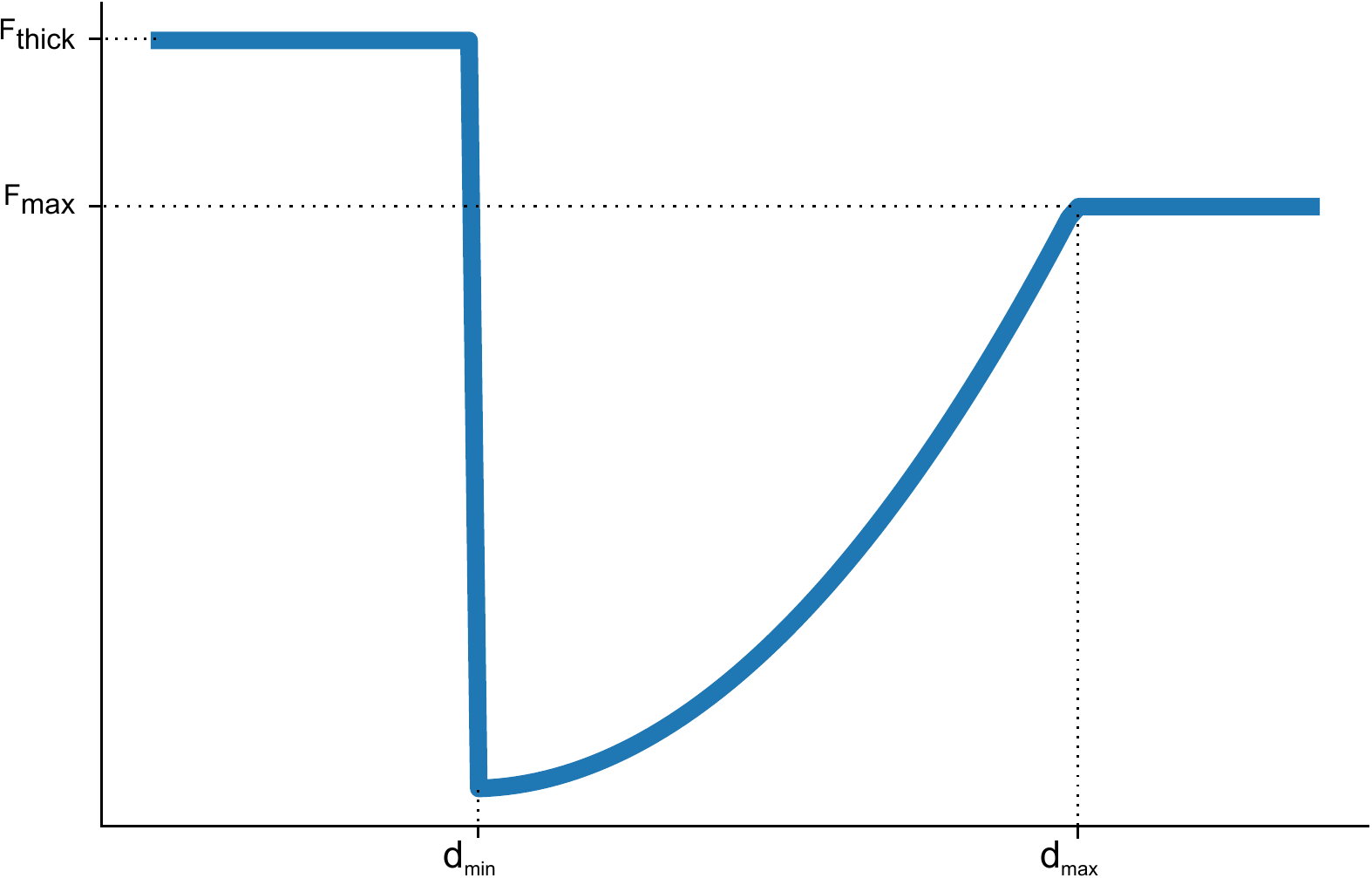}
\caption{Forbidden zone function $F(p)$. 
Because Zometool elements that stay close to $d_{\text{min}}$ {are preferred,
elements farther from $d_{\text{min}}$ are penalized.}
}
\label{fig:forbidden}
\end{figure}
\begin{description}[style=unboxed,leftmargin=0cm]
\item[Forbidden Zone]
\minorhl{
\mbox{$F(p)$} is defined as a quadratically increasing function that is dependent on the distance between the node n and the nearest triangle centroid on surface \mbox{$M$} when the distance is within the range from $d_{\text{min}}$ to $d_{\text{max}}$ (as shown in \mbox{\figname~\ref{fig:forbidden})}.
}
\myhl{
We choose appropriate parameters by conducting tests on different 3D shapes. We thus identify   the following set of parameters that produce elements that do not extrude out the forbidden zone across all the tested shapes.
}
\myhl{
$F(p)$ is set as (i) a constant large penalty ($F_{\text{thick}}$) when the distance is less than $d_{\text{min}}$ to prevent the Zometool elements from growing between the outer surface and $d_{\text{min}}$, and 
(ii) $F_{\text{max}}$ when the distance exceeds $d_{\text{max}}$. 
}
\myhl{
By testing various 3D shapes, we found that the following set of parameters produce results that do not extrude out of the forbidden zone for all the tested shapes:
}
$d_{\text{min}} = 16.0$ ($\frac{1}{3}$ length of $b_0$) , $d_{\text{max}} = 47.3$ (length of $b_0$), $F_{\text{max}}=70.0,$ and $F_{\text{thick}} = 90.0$; 
these values are thus used all examples in this paper. 
\end{description}

\subsubsection{Regularity}
\begin{figure}[ht]
\centering
\includegraphics[width=1.0\linewidth]{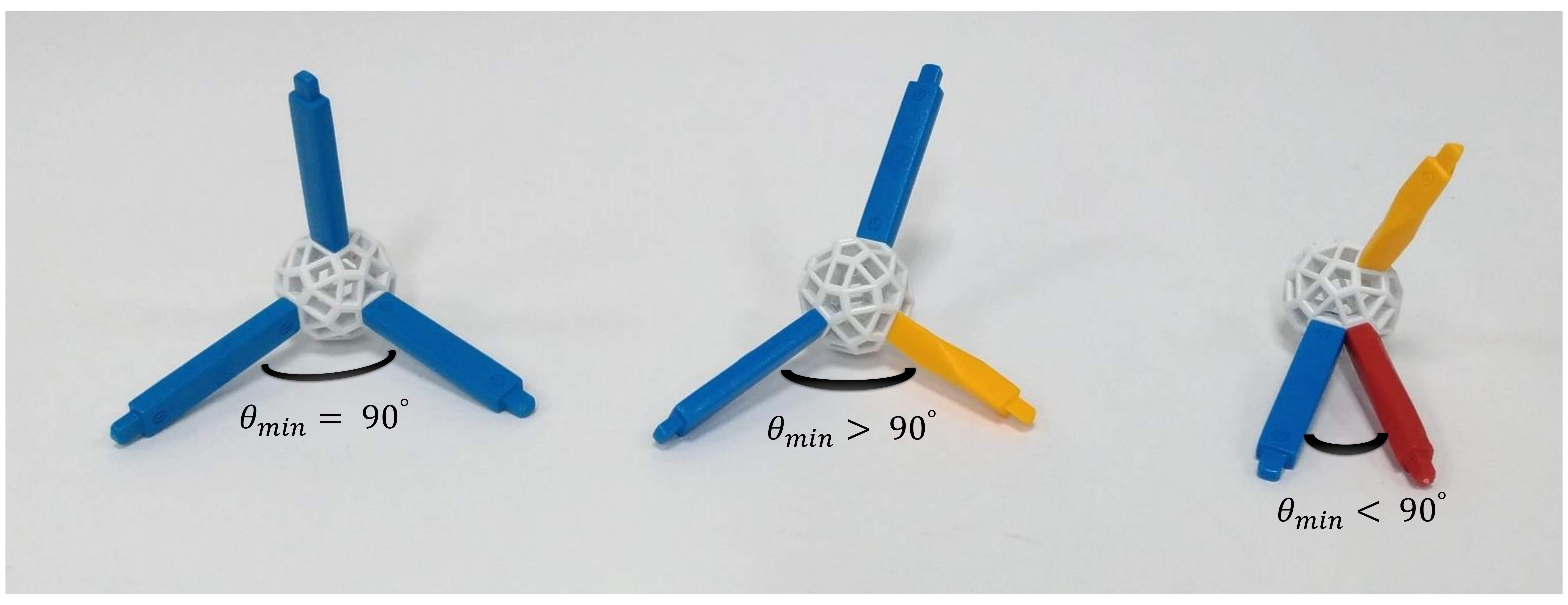} 
\caption{\textbf{Regularity.} We penalize configurations where the angle between the struts stray far from $90^\circ$.}
\label{fig:Regularity}
\end{figure}

\myhl{
On observing the assembling process of several complicated Zometool structures, we discover two major states that slow down the process: (i) the struts on a Zomeball are too close to each other (\ie, the angles between the struts are too small), and (ii) the slots used to connect the struts are spread out irregularly on the Zomeball.
}
To address these concerns, we regularize the angles between struts to be exactly $90^\circ$ and penalize any angle that is too small or too large (\figname~\ref{fig:Regularity})

\begin{align}
E_{\text{reg}}(\mathbf{Z}) = \sum_{i=1}^{S} \frac{1}{|\mathcal{N}_i|} \sum_{s_j\in\mathcal{N}_i} (\text{min}(\theta_{ij})-\frac{\pi}{2}),
\end{align}
\myhl{where \mbox{$S$} represents the number of struts} and $\mathcal{N}_i$ represents the struts adjacent to strut $i$.
The resulting  Zometool structure has a relatively high number of repeated Zomeball patterns (the struts on each Zomeball are placed at similar slots), and this considerably reduces the assembling time.

\subsubsection{Valence}
\begin{figure}[ht]
\centering
\includegraphics[width=1.0\linewidth]{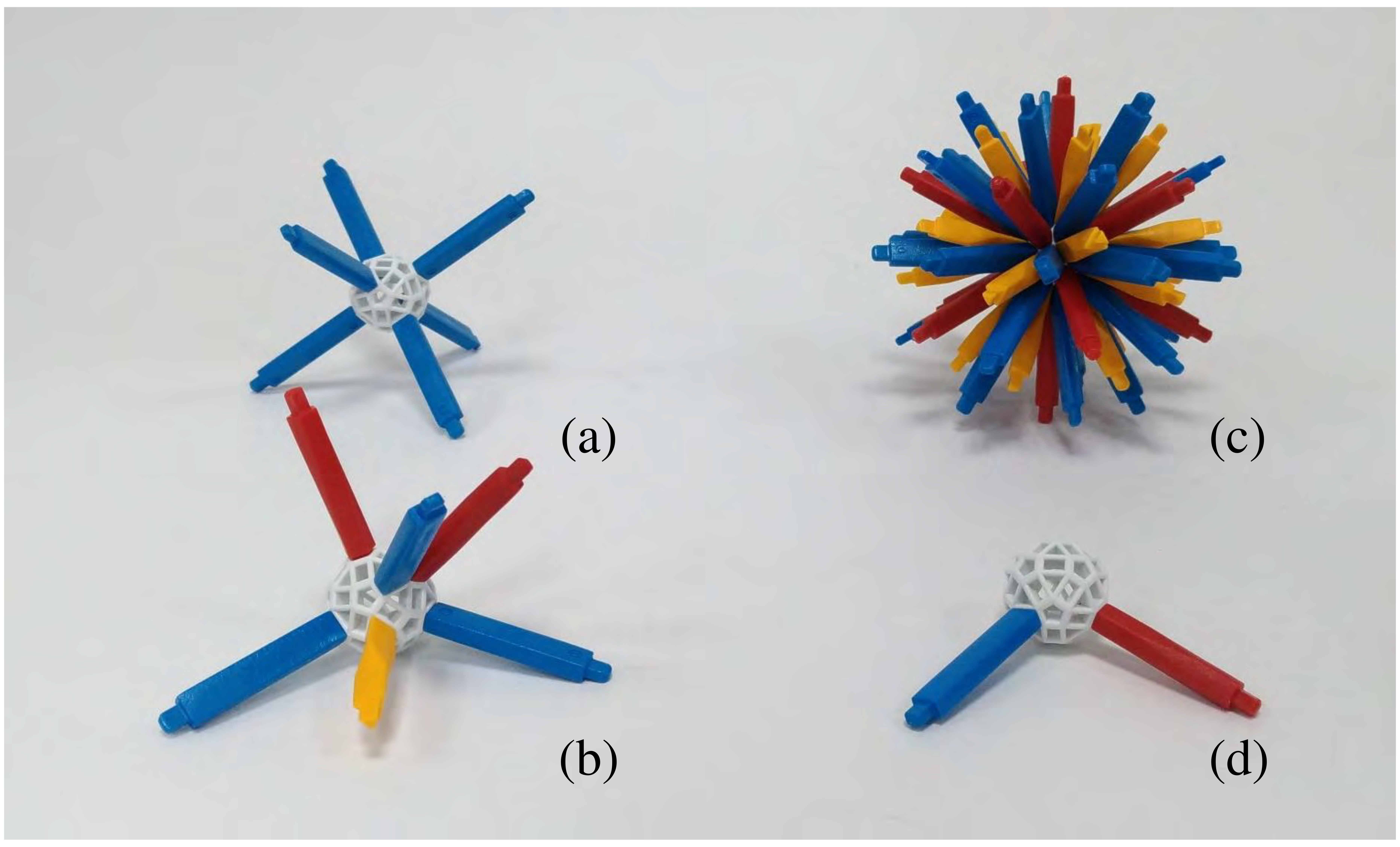} 
\caption{\textbf{Valence.} 
We encourage the valence of each Zometool node to be 6 (as in configuration (a) and (b)). 
We penalize the valence that is not 6 (configuration (c) and (d)).
}
\label{fig:Valence}
\end{figure}

For ease of assembly, we regularize the optimized Zometool structure such that it has good valence for simple structures (see \figname~\ref{fig:Valence}). 
\myhl{
The primary reason for regularizing valence number is that an undesired Zomeball valence number (\mbox{\figname~\ref{fig:Valence}} (c) and (d)) increases the possibility of element collision, which considerably reduces the optimization efficiency.
Collided structures are considered to be invalid (because they can not be assembled) in our method (this is discussed subsequently herein).
}
We set the target valence as $6$ according to the initial cube structure, which minimizes the complexity and maximizes the utility of each Zomeball:
\begin{align}
E_{\text{val}}(\mathbf{Z}) = \sum_{i=1}^{N_{in}} \frac{(V_i-6)^2}{6},
\end{align}
where \minorhl{\mbox{$|N_{in}|$} denotes the number of internal nodes and $V_i$ denotes the valence of node \mbox{$n_i \in N_{in}$}} (\figname~\ref{fig:Valence}). 

\subsubsection{Shape simplicity}
\myhl{
To reduce the complexity of the Zometool structure, we limit the number of Zometool elements.
However, oversimplification negates the effects of the shape fidelity energy discussed in \mbox{\secname~\ref{sec:opt_distance}; that is, the desired shape cannot be adequately replicated}.
For the optimal trade-off of shape fidelity and simplicity, we determine the target number of Zometool elements before optimization.
To this end, we run a random Zometool assembling process ten times within the volume of the input shape.
For each run, we record the number of Zometool elements required to reach a predefined shape similarity threshold between the Zometool structure and the target shape $M$. 
We set the 90\% of the average number of Zometool elements as the final target item number ($\tau$).
For most cases, this target number serves as the upper bound of the number of elements, and the simulated annealing optimization process is always terminated before this target number is reached.
}
\minorhl{Let \mbox{$|N|$} represent the total number of nodes, and \mbox{$|S|$} represent the total number of struts.
}
Simplicity is measured as the difference between the actual and target item numbers:
\begin{align}
\eqhl{E_{\text{sim}}(\mathbf{Z}) = \frac{1}{\tau}(|N|+|S|-\tau)^2.}
\end{align}

\subsection{Exploration Mechanism}
Determining the Zometool structure that minimizes the energy $E(\mathbf{Z})$ (\eqname~\ref{eq:sa_energy}) is a \chireplace{non trivial}{nontrivial} optimization problem because $E(\mathbf{Z})$ is \chireplace{non convex}{nonconvex} and contains global terms. 
Because an exhaustive search is impractical, we adopt a more scalable strategy based on the simulated annealing algorithm~\cite{Salamon:2002:SA}.
In a nutshell, this algorithm executes a random exploration of the solution space by iteratively perturbing the current solution with a certain probability depending on the energy variation between the two solutions and a relaxation parameter $T$.
We describe our local perturbation operators and relaxation scheme as follows.
Algorithm~\ref{alg:exploration} details our optimization algorithm.

\begin{algorithm}[!ht]
\caption{Exploration mechanism}
\label{alg:exploration}
\begin{algorithmic}[1]
\Input{Initialized Zometools $\bar{\mathbf{Z}}$,\\ relaxation parameter $T=T_{init}$}
\Output{Optimized Zometoos $\mathbf{Z}$}
\Procedure{Exploration}{$\mathbf{Z}$}
\Repeat
    \State generate $\mathbf{Z}'$ from $\mathbf{Z}$ with a random local operation.
    \State draw a random value $p \in [0, 1]$ 
    \If{$p < \text{exp}(\frac{E(\mathbf{Z})-E(\mathbf{Z}')}{T})$ and CollisionFree($\mathbf{Z}$)} 
        \State update $\mathbf{Z}' \leftarrow \mathbf{Z}$
    \EndIf
    \State Update $T \leftarrow C\times T$ \Comment{Update temperature.}
\Until{$T< T_{end}$}
\EndProcedure
\end{algorithmic}
\end{algorithm}

\begin{figure}[ht]
\centering
\includegraphics[width=1.0\linewidth]{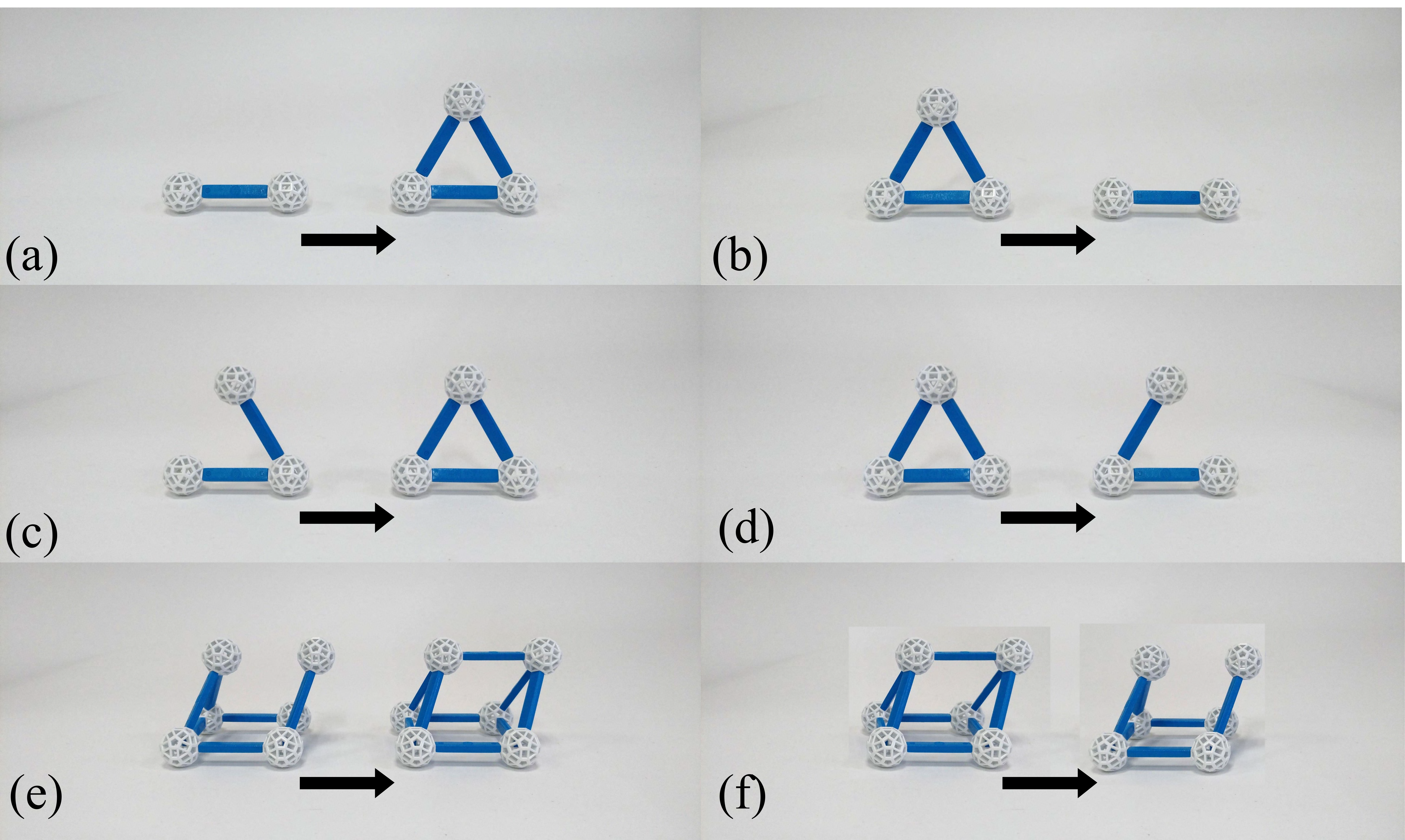} 
\caption{
We use six local operations during structure perturbation. (a) InsNode, (b) DelNode, (c) InsStrut, (d) DelStrut, (e) InsBridge, and (f) DelBridge.
Each operation is performed from the left configuration to the right configuration.
}
\label{fig:local_op}
\end{figure}

\begin{description}[nosep,itemsep=0pt,leftmargin=0pt]
\item[Local Perturbation Operation]
During the exploration, we proposed \myhl{six} local perturbation operations (\figname~\ref{fig:local_op}) to construct the Zometool structure by minimizing \eqname~\ref{eq:sa_energy}.
\begin{itemize} 
\item \textbf{InsNode} This operator inserts a new node and  two struts to split the original strut.
\item \textbf{DelNode} This operator deletes a node and two struts.
\item \textbf{InsStrut} \minorhl{This operator inserts a single strut to connect two nodes that are not directly linked.}
\item \textbf{DelStrut} \minorhl{This operator deletes a strut between two nodes that are directly linked.}
\item \textbf{InsBridge} This operator inserts a new strut to merge two \myhl{nodes that are not directly linked (two nodes with connected path length more than two)}. 
\item \textbf{DelBridge} This operator deletes a strut between two \myhl{nodes with connected path length more than two}. 
\end{itemize}
\end{description}

\emph{Operation validity}
The chosen operation at each iteration updates the structure from the previous iteration. 
However, some updates might introduce invalid structures, such as structures with item collision.
We detect collision by checking the (1) Zomeball to Zomeball, (2) strut to Zomeball, and (3) strut to strut distances;
if the distance is less than a certain threshold, the update is rejected.

\subsubsection{Cooling schedule}
The relaxation parameter $T$, referred to as temperature, controls both the speed and quality of the exploration.
Starting from \chinky{an} initial temperature $T_{init}$, we decrease the temperature, approaching  zero as the iteration tends to infinity.
This process is referred to as cooling, and various cooling schedules are available.
Although the logarithmic cooling schedule~\cite{Salamon:2002:SA} guarantees global minimum convergence, we implement a geometric cooling schedule~\cite{Henderson:2003:SA}. 
In our experiment, we set the initial temperature $T_\text{init} = 1$ and apply the decrease rate $C=0.99$ after every $100$ iterations.

\section{Object Partition}
\label{sec:surf_part}
\begin{figure}[h!]
\centering
\includegraphics[width=1.0\linewidth]{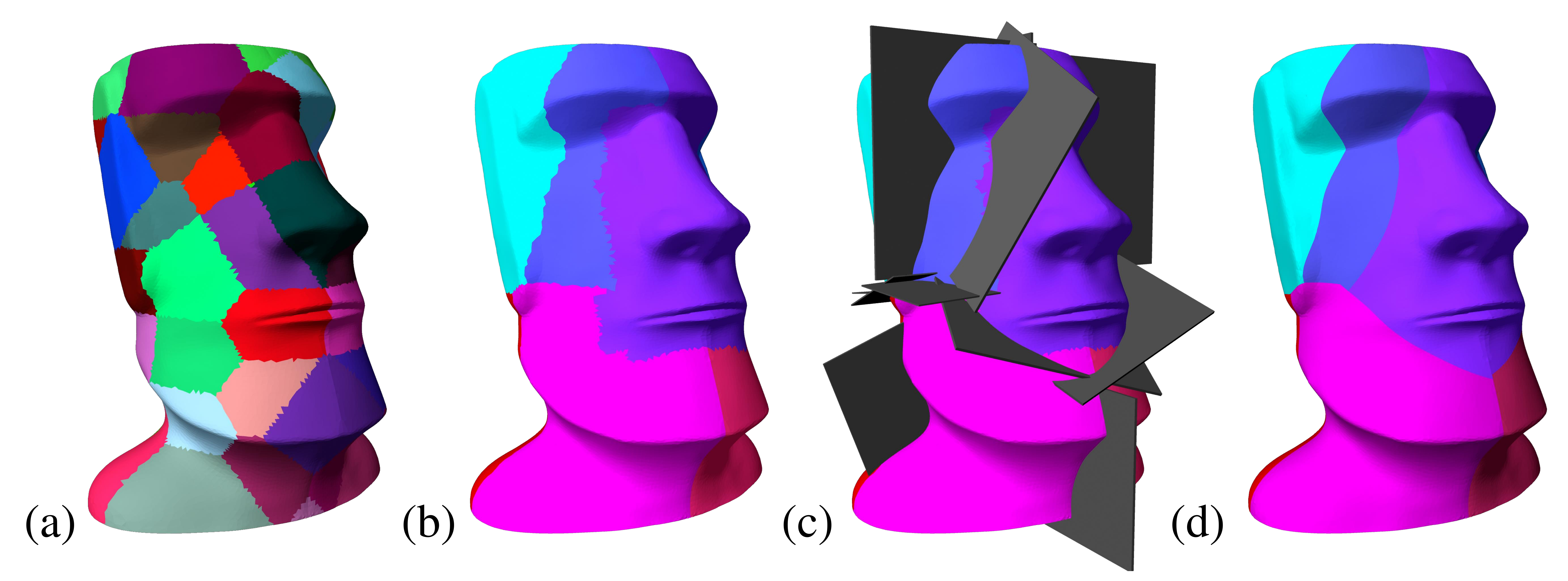} 
\caption{(a) Result of nearest node classification (b) result of graph cut, (c) cut-plane generated using an SVM classifier, and (d) cutting result.
} 
\label{fig:cut_plane}
\end{figure}

Most consumer-level 3D printers have limited printing volumes.
To print large-scale objects, the objects  must be decomposed into smaller partitions.
Conventional surface partition methods only account for surface features; however, we must also consider the relationship between the outer surface partitions and the inner optimized Zometool structure.
Specifically, in this work, we place connectors between the outer and inner structures in order to connect them (see~\secname~\ref{sec:fab}). 
We can simply compute the distance between each triangle $t$ and all nodes in $\mathbf{Z}$ and assign $t$ to the nearest node as its label. 
However, inconsistencies may arise among adjacent triangles, engendering unsatisfactory visual effects and assembly complexities because numerous small partitions might exist (\figname~\ref{fig:cut_plane}(a)).


To address this concern, we formulate the problem as a multi-label graph-cut minimization problem.
As each triangle $t$ can potentially correspond to different Zometool node\chinky{s}, it gets assigned data costs for different corresponding nodes.
Given $n$ elements, $k$ labels, and $n\cdot k$ costs, finding the minimum assignment is a typical NP-hard combinatorial problem, which we solve using
Boykov~\cite{boykov:2004:experimental}.

After partitioning the structure, we further regularize the boundaries between the partitions by performing multi-class classification;
this process of seeking smooth boundaries ensures easy assembly.
\myhl{This problem has been previously addressed by \mbox{Wang~\etal~\cite{wang2016improved}} and \mbox{Alderighi~\etal~\cite{Alderighi:2018:MCD}}.}
We follow \cite{wang2016improved} and use an SVM~\cite{cortes1995support} classifier to determine our cut plane. 
We describe the formulation and implementation of a Markov random field (MRF) problem and how to find the cut planes in the following paragraph.

\subsection{Surface Partition}
\subsubsection{Optimization energy}
We compute the assignment function $f$ that assign labels to each triangle $t$, where $t \in T$, such that the labeling $f$ minimize the following energy $E(f)$:
\begin{align} \label{eq:graph}
E(f) = w_{data} \sum_{t\in T}D(t, f_t) + w_{smoothness} \sum_{t,s\in \mathcal{N}} \psi_{t,s}(t, s, f_t, f_s),
\end{align}
where \myhl{\mbox{$f_t$ and $f_s$ are labels assigned to triangle $t$ and $s$}, and $\mathcal{N}$ is the set of all pairs of triangles sharing edges.}
We set $w_{data}=1.0$ for all the shapes shown in this paper,
and we use \mbox{$w_{smoothness}$} to control the size of the partitions, which
is crucial for limiting the size of the partitions such that they fit within the 3D printing volume.
Specifically, we initialize \mbox{$w_{smoothness}$} as 10 and examine whether we can fit all the partitions within the 3D printing volume. 
If not, \minorhl{we multiply \mbox{$w_{smoothness}$} by 0.1} and repeat this process until all the partitions fit into the desired volume.
We optimize this function by using the multilabel graph-cut algorithm proposed by Boykov~\cite{boykov:2004:experimental}.
In our setting, the entire outer nodes of $\mathcal{Z}$ form the complete possible label set $L$.
This $E(f)$ comprises two terms: data cost and smoothness.

\emph{\textbf{Data cost.}}
\minorhl{We measure how well a triangle $t$ covers an outermost node $n \in N_{out}$ as data cost.}
\minorhl{
This cost is simply defined as the distance between the centroid of the triangle $t$ and the outermost node $n \in N_{out}$.}
\begin{align}
\eqhl{D(t,f_t) = \log (d(t, n)).}
\end{align}
\minorhl{
Noted that the set of possible $f_t$ is the set of node id of outermost nodes.
}

\begin{figure}[h!]
\centering
\includegraphics[width=1.0\linewidth]{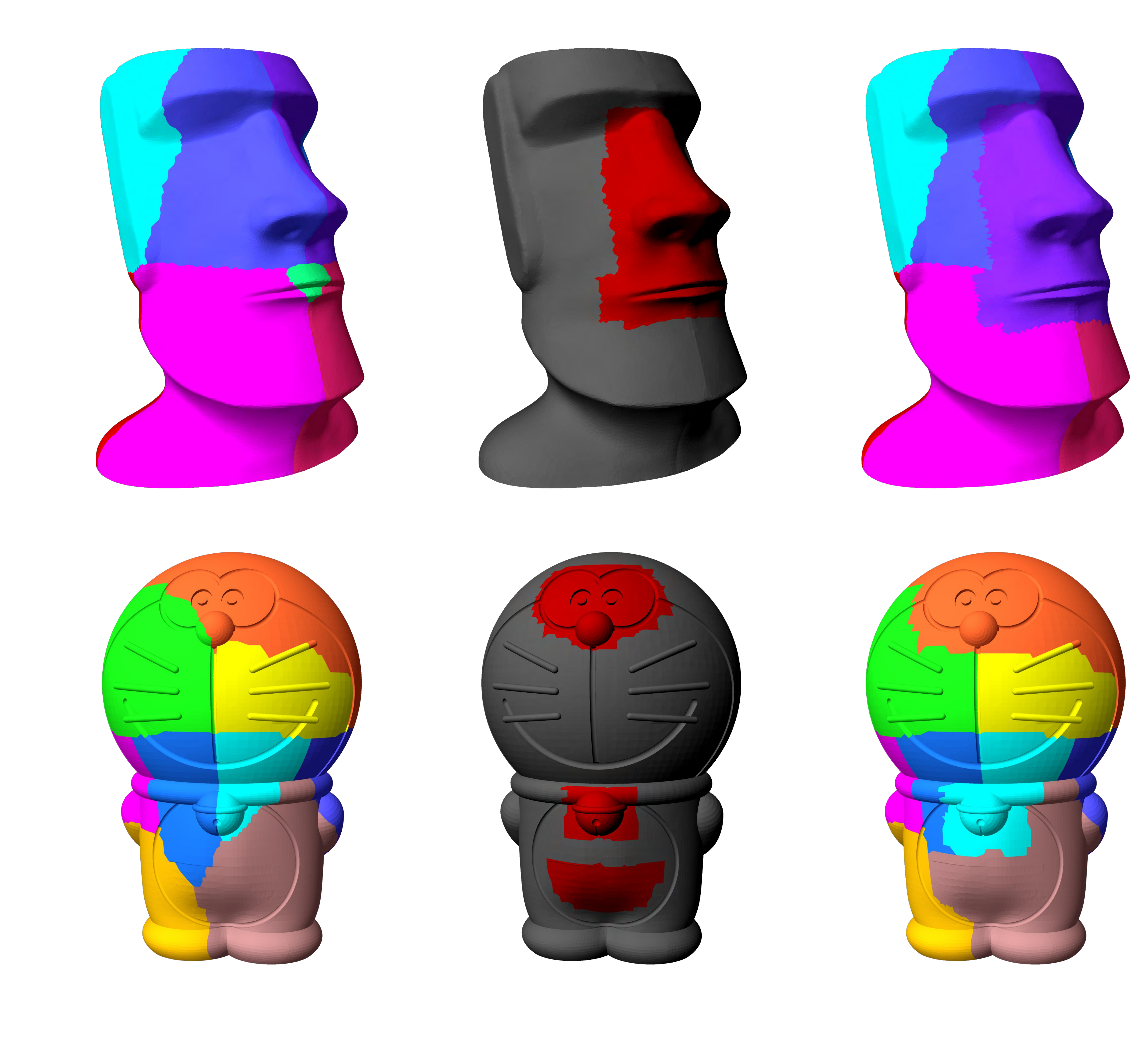} 
\caption{
In the left-most column, we show optimized partitions without introducing the user-guided saliency term.
We can observe that the partition cut through some salient features on the mesh.
To address this problem, the user annotates the region that they wish to preserve (red region in the middle column).
We then incorporate an additional term in the original smoothness term to prevent the annotated regions from separating into multiple parts (right-most column).} 
\label{fig:saliency}
\end{figure}

\emph{\textbf{Smoothness cost.}}
This term measures the spatial consistency of neighboring elements.
\begin{align}
\psi_{t,s}(t, s, l_t, l_s) = 
\begin{cases}
0, & \text{if } l_t = l_s, \\
-\log(\theta_{t,s}/\pi)\varphi_{t,s} + w_{saliency} * E_{saliency}(t), & \text{otherwise} 
\end{cases}
\end{align}
where we set $w_{saliency}=5.0$, and $\theta_{p,q}$ and $\varphi_{p,q}$ are the dihedral angle and the centroid distance between triangles $p$ and $q$, respectively.
With the smoothness term, two adjacent triangles are likely to have consistent labels.
Each object has many salient regions that the  the partition seam must not go through (e.g., the eyes and nose on the face \chireplace{~\figname~\ref{fig:saliency}}{in \figname~\ref{fig:saliency})}.
\chireplace{In order to}{To} preserve the integrity of the salient regions, we ask users to draw the region that they wish to preserve, and we formulate this requirement as part of the smoothness cost to prevent the \chireplace{the partition seam}{partition seams} from cutting through the salient region (see last column of \figname~\ref{fig:saliency}).

\begin{align}
E_{saliency}(t) = 
\begin{cases}
1, & \text{triangle is marked as salient region}, \\
0, & \text{otherwise} 
\end{cases}
\end{align}

\subsubsection{Optimized partitions}
We solve the labeling problem~in \eqname~\ref{eq:graph} by using a graph-cut algorithm and obtain 11 labels on the Maoi head object~(\figname~\ref{fig:cut_plane}(b)).
Compared with the method that directly assigned the triangle to the closest outermost node,
our approach yields \chinky{much} smaller partition numbers (11 vs. 60 (\figname~\ref{fig:cut_plane}(a)), and each partition is of relatively better shape and size.

\subsection{Object Cut}
To cut the physical object into pieces, we must find the cut planes that separate the space occupied by the object.
However, the boundaries between the optimized partitions are not regular enough to be directly used as the separating plane (see \figname~\ref{fig:cut_plane}(b) and the inset).
It is very difficult to find the plane in 3{D} space because of the varied plane normal vectors.
\begin{wrapfigure}{r}{0.15\columnwidth} 
\includegraphics[width=0.15\columnwidth]{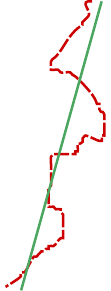}
\end{wrapfigure}
Moreover, the zig-zag partition boundaries introduce additional complexities in the 3D printing process; thus, we find a clean separating plane
by following \cite{wang2016improved} and analogizing our problem as a multiclass classification problem, which is then solved using SVM.
We use each triangle as a data sample, with it's location as \chinky{the} feature vector and the optimized label as it's class.
Briefly, the SVM algorithm finds the best hyperplane, which is the one that represents the largest separation, or margin, between the two adjacent classes.
\myhl{
Because we found that the hyperplanes obtained using SVM suffice to cut the planes, we did not proceed with the subsequent step suggested described in \mbox{\cite{wang2016improved}} (\mbox{\figname~\ref{fig:cut_plane}(c)}).
}


\section{Fabrication}
\label{sec:fab}
The object to be printed should be a solid mesh.
In our method, we use the Zometool structure to fill in most of the inner volume while retaining \myhl{a predefined thickness}\ignore{the minimum thickness} of the outer shell.
The minimum thickness is dependent on the  requirements of the 3D printer used.
To print each solid piece, we must obtain the inner surface and use the original surface as the outer surface.
The inner surface can be generated in many approaches, such as by shrinking the mesh along the vertex normals.
However, these approaches often generate surfaces with flipped triangles that stick out of the outer surface.
\minorhl{
To prevent this problem, we instead voxelize the original object and remove the voxels whose center falls within the minimum thickness from surface.
}
We then use the outermost surface of the remaining voxels as our inner surface solid piece.
\myhl{
As discussed in \mbox{\secname~\ref{sec:opt_distance}}, we penalize the Zometool elements that grow within the range of $d_{\text{min}}$ (which is our default thickness).
For all the shapes shown in this paper, no Zometool elements poked out beyond the surface.
}

\subsection{Generate connector}
\begin{figure}[ht]
\centering
\includegraphics[width=1.0\linewidth]{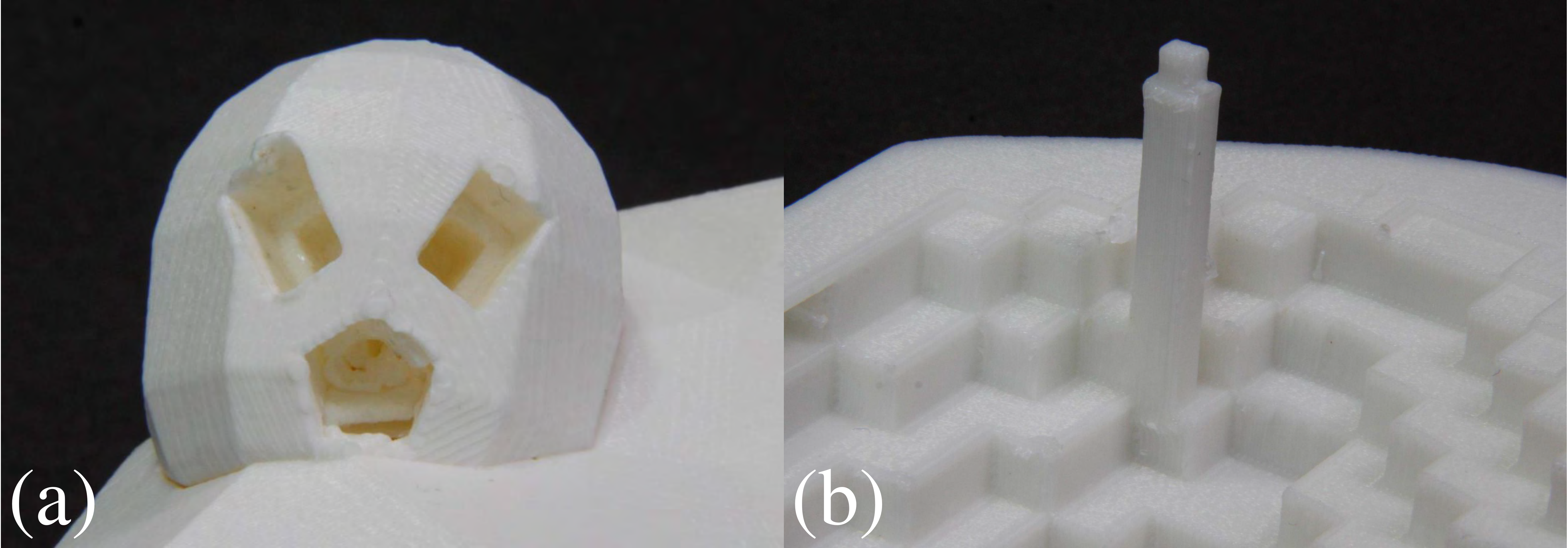} 
\caption{
Connector designs: (a) dig holes and (b) grow tenons on inner surface.
The materials in the dug holes can not be removed entirely in (a), so the struts can not be inserted well.
}
\label{fig:exp_connector}
\end{figure}

With the inner surface, we need to place connectors on it to connect between inter Zometool structure and outer shells.
Two potential designs for building these connectors are:
\begin{enumerate}
\item Dig holes on the surface and use the Zometool struts to connect both inter and outer structures (\figname~\ref{fig:exp_connector} (a)).
\item  Grow Zometool tenons on the inner surface (\figname~\ref{fig:exp_connector} (b)).
\end{enumerate}
We tried both designs, and as we experimented, we observed that the generated support structures from the 3D printers hugely reduce the quality of the digged holes.
The reason is the printed holes are usually filled with the support materials, and it is difficult to remove all of them.
Hence, we choose to grown tenons on the inner surface with following method.
    
\emph{\textbf{Grow tenons on the surface.}}
Given it's easier cleanup and more robust structure, we use this design to connect the inner Zometool structure and outer shell (see \figname~\ref{fig:exp_connector} (b)). 
\myhl{We restricted the grown tenons have to be perpendicular to the inner surface in order to achieve better structural robustness. }
And we decide how many tenons on each outer shell with the following process:
\minorhl{We shoot rays from each admissible slot on a single Zomeball in the optimized Zometool structure, and record whether it intersects with the tested shell.
Since the tenons are only generated perpendicular to the inner voxel surface, it's direction is basically axis-aligned. 
This means that for each Zomeball, only 6 slots are admissible for each tenon.
We repeated this examination on all of 6 slots on each of the Zomeball covered by this shell, and we find the directions that are perpendicular to the inner surface and grow tenons along those directions.
}

    
\section{Result}
\label{sec:result}
\subsection{Experiment environment}
We implement ZomeFab in C++ and Python on desktop PC with 3.4GHz CPU and 16GB memory. 
The computation time of Zometool structure usually takes around 0.5 - 2.5 hrs due to the convergence of the simulated annealing. 
The multi-label graph cut algorithm takes around 2 mins for all shapes shown in the paper.
The outer surfaces are printed by Ultimaker 3, a low-cost FDM 3{D} printer with 0.2m x 0.2m x 0.2m printing volume and PLA material. 

\subsection{Evaluation}
We evaluate the material cost and fabrication time between ``Zomefab" and solid-printed object (denoted as ``baseline" method).
We use CURA\footnote{https://ultimaker.com/en/products/ultimaker-cura-software}, a slicer software for 3{D} printing, to simulate the fabrication time and used materials for solid-printed object.
Meanwhile, we subdivide the shape using a octree, and stop the subdivision once each sub-shape fit into the printing volume.
To evaluate our method, we report two different infill methods called ``hollow" and ``solid".
Shapes printed under ``hollow" use only 20\% infill rate, where ``solid" use 100\% infill rate.
Note that we only use 20\% infill rate to fabricate the results shown in this paper, and the ``solid" results are simulated using CURA.


\emph{\textbf{Material cost.}}
As shown in \tabname~\ref{tab:result_cost}, our method greatly saves materials from 24\% to 64\% under ``hollow'' setting, and from 68\% to 85\% under ``solid'' setting.
As expected that our method brings more benefits when infill rate grows.
The material cost are listed as follow: 0.56 USD/meter, Zometool strut: 0.19 USD/strut and Zomeball: 0.29 USD/ball for our experiments. 

\emph{\textbf{Fabrication time.}}
We report printing time evaluation under single 3D printer scenario and assembling time of zometool structure in \tabname~\ref{tab:result_cost}.
The statistics shows that the overall fabrication time, \ie~printing time of outer pieces plus the assemble time of Zometool structure, is on average $30\%$ shorter than the baseline method under ``hollow'' setting, and $74\%$ under ``solid'' setting.

\subsection{Zometool Use}
Table \ref{tab:result_Zometool} shows the quantity of Zometool struts and balls used in each result. 
In \secname~\ref{sec:Zometool}, we use smallest blue struts to make the zomecube as the unit structure in order to make the best-fit initial structure. 
The bigger the object is, more smallest blue struts (${b_0}$) will be used.
We also observed that, blue struts and yellow struts are more likely to be used interchangeably since we encourage regularity during simulated annealing optimization. 
Meanwhile, the rule of Zometool indicates that the longest strut only can be replaced by one shortest and middle strut in the same color.
So it seldom introduce red struts into the structure, which resulted in the used number of struts imbalance shown in \tabname~\ref{tab:result_Zometool}.
\myhl{To ease the assembling process of Zometool structure, we wrote a simple interface that can highlight the connected struts for each ball.
This can greatly improve the efficiency of the assembling process.
}

 \begin{table*}[ht]
\centering
\resizebox{1.\linewidth}{!} {
\begin{tabular}{|c|c|c|c c c| c c c|c | c|} \hline
 \multirow{2}*{Mesh} & \multirow{2}*{Infill Method} & \multirow{2}*{Fabrication Method} & \multicolumn{3}{c|}{Single 3{D} Printer Fabrication Time (hours)} & \multicolumn{3}{c|}{Material Cost (USD)} & \multicolumn{2}{c|}{Efficiency (Saved)}\\\cline{4-11} 
 & & & 3D printing & Zometool (assemble time) & Overall (sum) & 3D printing & Zometool & Overall (max) & Time & Material\\ \hline
 
 \multirow{4}*{Moai} & \multirow{2}*{Hollow} & Zomefab & 293.97 & 2.5 & 296.47 & 82.56 & 70.57 & 153.13 & \multirow{2}*{31.70\%} & \multirow{2}*{13.83\%}\\ 
 &  & Baseline & 434.08 &  & 434.08 & 177.71 &  & 177.71 & &\\\cline{2-11}
 & \multirow{2}*{Solid} & Zomefab & 523.20 & 2.5 & 525.70 & 170.9 & 70.57 & 241.47 & \multirow{2}*{77.75\%} & \multirow{2}*{65.11\%}\\
 &  & Baseline & 2362.28 & & 2362.28 & 692.07 & & 692.07 & &\\ \hline
  
 \multirow{4}*{Squirrel} & \multirow{2}*{Hollow} & Zomefab & 355.05 & 3.0 & 358.05 & 100.38 & 77.55 & 177.93 & \multirow{2}*{22.04\%} & \multirow{2}*{21.74\%}\\ 
 &  & Baseline & 459.28 & & 459.28 & 227.37 & & 227.37 & &\\\cline{2-11}
 & \multirow{2}*{Solid} & Zomefab & 643.65 & 3.0 & 646.65 & 213.96 & 77.55 & 291.51 & \multirow{2}*{76.57\%} & \multirow{2}*{66.17\%}\\
 &  & Baseline & 2759.72 & & 2759.72 & 861.66 & & 861.66 & &\\ \hline
 
 \multirow{4}*{Doraemon} & \multirow{2}*{Hollow} & Zomefab & 356.60 & 4.0 & 360.60 & 101.38 & 76.34 & 177.72 & \multirow{2}*{37.88\%} & \multirow{2}*{21.91\%}\\ 
 &  & Baseline & 580.48 & & 580.48 & 227.59 & & 227.59 & &\\\cline{2-11}
 & \multirow{2}*{Solid} & Zomefab & 643.65 & 4.0 & 647.65 & 227.78 & 76.34 & 304.12 & \multirow{2}*{57.71\%} & \multirow{2}*{64.25\%}\\
 &  & Baseline & 1531.42 & & 1531.42 & 850.79 & & 850.79 & &\\ \hline
 
\multirow{4}*{Totoro} & \multirow{2}*{Hollow} & Zomefab & 273.68 & 3.0 & 276.68 & 72.32 & 61.48 & 133.80 & \multirow{2}*{27.76\%} & \multirow{2}*{36.38\%}\\ 
 &  & Baseline & 383.02 & & 383.02 & 178.88 & & 178.88 & &\\\cline{2-11}
 & \multirow{2}*{Solid} & Zomefab & 468.82 & 3.0 & 471.82 & 148.26 & 61.48 & 209.74 & \multirow{2}*{77.68\%} & \multirow{2}*{68.09\%}\\
 &  & Baseline & 2113.42 & & 2113.42 & 657.31 & & 657.31 & &\\ \hline
 
\multirow{4}*{Iron Man} & \multirow{2}*{Hollow} & Zomefab & 282.30 & 2.0 & 284.30 & 73.99 & 57.76 & 131.75 & \multirow{2}*{34.98\%} & \multirow{2}*{34.12\%}\\ 
 &  & Baseline & 437.25 & & 437.25 & 199.97 & & 199.97 & &\\\cline{2-11}
 & \multirow{2}*{Solid} & Zomefab & 477.03 & 2.0 & 479.03 & 149.67 & 57.76 & 207.43 & \multirow{2}*{80.65\%} & \multirow{2}*{72.93\%}\\
 &  & Baseline & 2475.22 & & 2475.22 & 766.24 & & 766.24 & &\\ \hline
 
\end{tabular}
}
\caption{ZomeFab's performance on saving time \& material time as compared to a baseline method.}
\label{tab:result_cost}
\end{table*}


\begin{table*}[ht]
\centering
\resizebox{0.7\linewidth}{!} {
\begin{tabular}{|c|c|c|c|c|c|c|c|c|c|c|c|} \hline
\multirow{3}*{Mesh} & \multicolumn{11}{c|}{Zometool} \\\cline{2-12} 
& \multicolumn{3}{c|}{Blue} & \multicolumn{3}{c|}{Red} & \multicolumn{3}{c|}{Yellow} & \multirow{2}*{Total struts} & \multirow{2}*{Total balls}  \\\cline{2-10} 
& S & M & L & S & M & L & S & M & L & & \\ \hline
Moai & 112 & 0 & 0 & 0 & 0 & 0 & 140 & 0 & 0 & 252 & 73  \\ \hline
Squirrel & 144 & 22 & 0 & 8 & 4 & 0 & 119 & 3 & 1 & 301 & 85  \\ \hline
Doraemon & 143 & 26 & 1 & 2 & 5 & 1 & 89 & 1 & 1 & 569 & 87 \\ \hline
Totoro & 95 & 0 & 0 & 0 & 0 & 0 & 137 & 0 & 0 & 232 & 60 \\ \hline
Iron Man & 93 & 0 & 0 & 0 & 0 & 0 & 124 & 0 & 0 & 217 & 57  \\ \hline
Owl & 115 & 0 & 0 & 0 & 0 & 0 & 159 & 0 & 0 & 274 & 70  \\ \hline
Pig & 61 & 12 & 0 & 10 & 8 & 0 & 53 & 6 & 0 & 150 & 47  \\ \hline
Slime & 132 & 0 & 0 & 0 & 0 & 0 & 182 & 0 & 0 & 314 & 80  \\ \hline
Lion & 78 & 0 & 0 & 0 & 0 & 0 & 101 & 0 & 0 & 179 & 50 \\ \hline
Bunny & 215 & 0 & 0 & 0 & 0 & 0 & 266 & 0 & 0 & 481 & 126  \\ \hline
\end{tabular}
}
\caption{Zometool element usage.
}
\label{tab:result_Zometool}
\end{table*}

 \begin{table*}[ht]
 \centering
 \resizebox{0.86\linewidth}{!} {
 \begin{tabular}{c} 
 \includegraphics{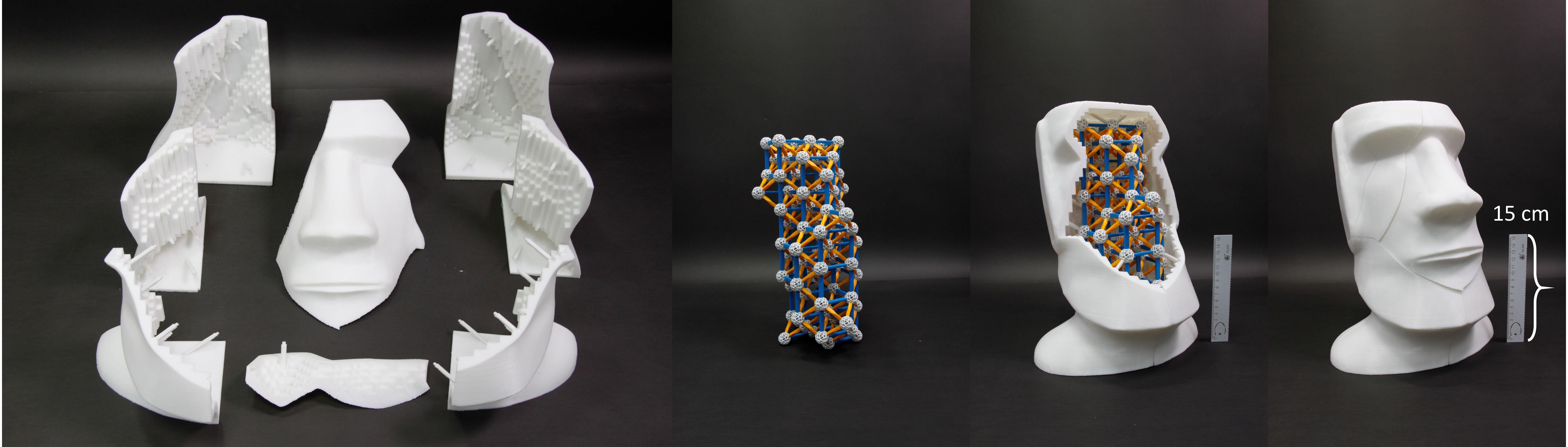} \\
 \includegraphics{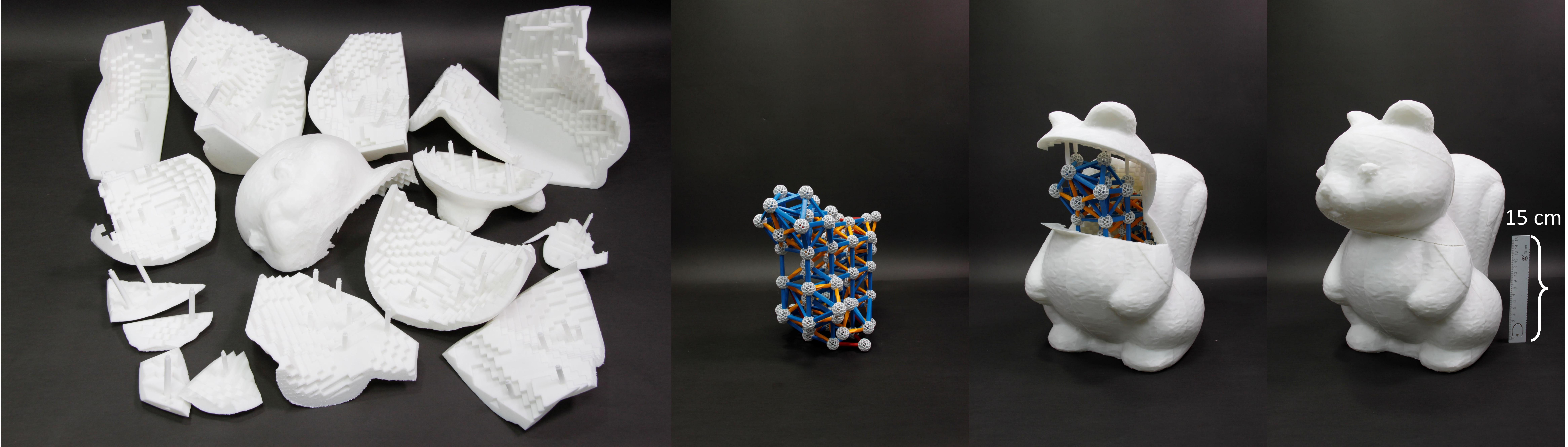}\\
 \includegraphics{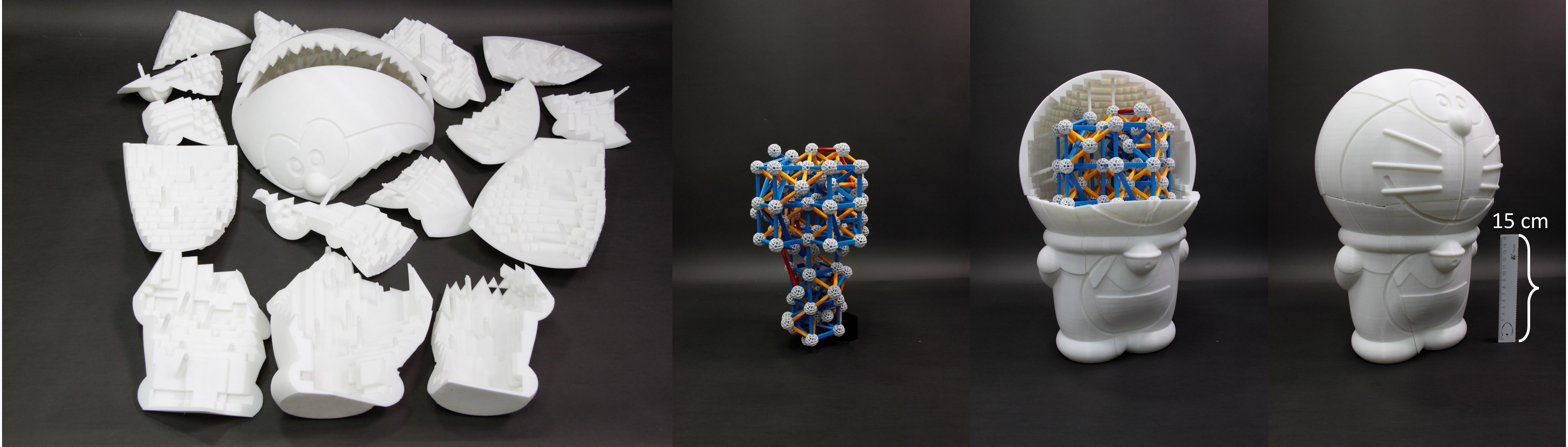}\\
 \includegraphics{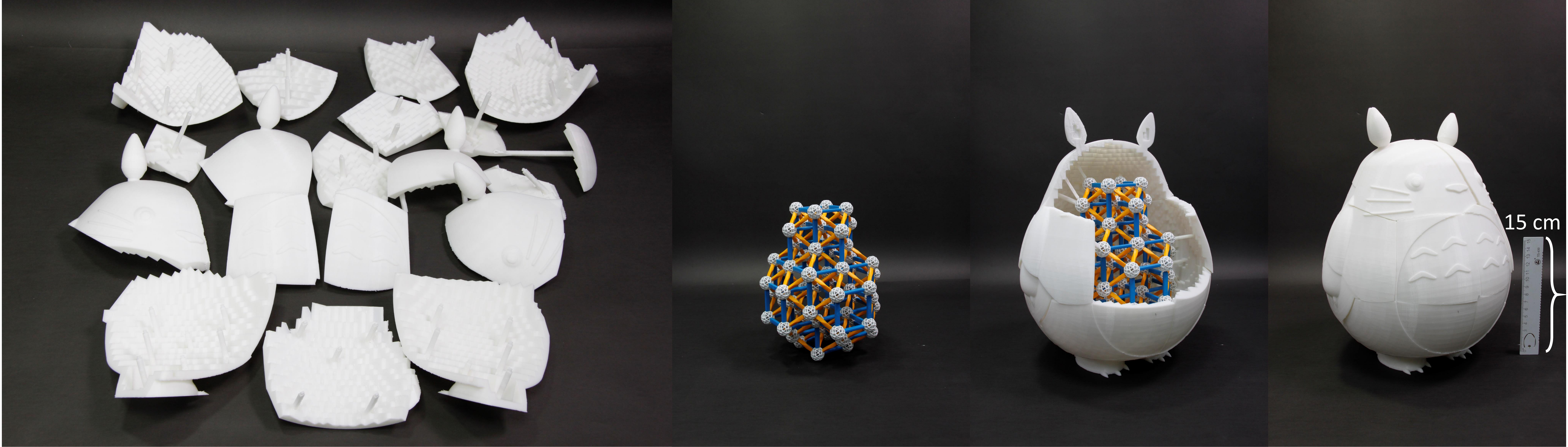}\\
 \includegraphics{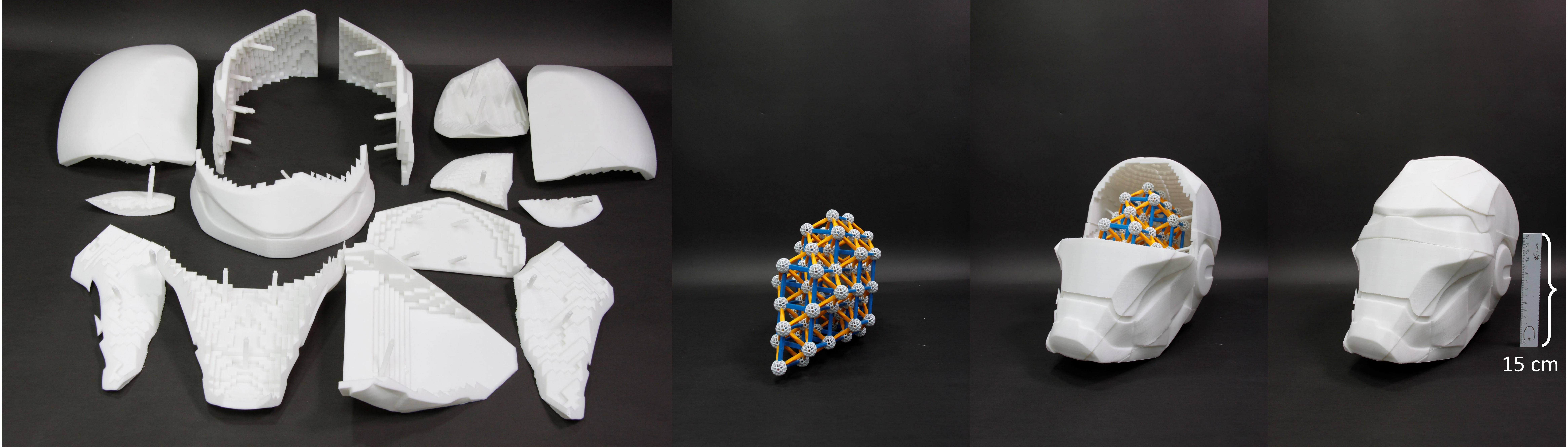}\\
 \end{tabular}
 }
 \caption{We fabricate\robin{d five} shapes using our method\robin{, and}
 show the printed pieces (1st column), inner Zometool structu\robin{r}e (2nd column), and assembled final results (3rd and 4th column).
 Noted that the scale of 2nd, 3rd, and 4th columns are the same. 
 }
 \label{tab:result_ZomeFab_real}
 \end{table*}

\section{Conclusion}
\label{sec:conclusion}
In this paper, we propose ZomeFab, a hybrid fabrication method that combines Zometool 
and 3{D} printing to fabricate a large-scale 3{D} object. 
We aimed at decomposing an input 3D object into a inner structure and pieces of outer surface. 
By replacing the huge inner space with Zometool from 3D printed materials, we can greatly reduce the cost of 3D printing.
And we retain the fine geometric detail by printing the outer shell with 3D printer.
With the reusability of Zometool, the long-term cost of fabrication is greatly decreased. 
We have demonstrated that our method is able to fabricate large-scale object by physically replicate 5 objects.

\emph{\textbf{Limitations and Future Work.}}
There are many remaining challenges and opportunities for future research on large-scale fabrication.
While our initial unit structure is simple and easy-to-assemble, it's size (4.7cm x 4.7cm x 4.7cm) prevents us to fabricate 3D object with thin structure.
Meanwhile, since Zometool is designed under formal and strict mathematical formulation, the connection between two separated Zometool sub-structure might not exist within the 3D object inner space.
As a result, our method can not fabricate 3D object with two or more parts connected by thin structure that our initial unit structure can not fit in.
In the future, we would love to investigate how to shrink the size of the initial unit structure, in order to enable more complicated large-scale 3D object fabrication.

\bibliographystyle{eg-alpha-doi}
\bibliography{zomeFab_pg}

\newcommand{\etalchar}[1]{$^{#1}$}
\begin{thebibliography}{\uppercase{BWBSH14}}

\bibitem[ACP{\etalchar{*}}14]{alemanno2014interlocking}
\textsc{Alemanno G., Cignoni P., Pietroni N., Ponchio F., Scopigno R.}:
\newblock Interlocking pieces for printing tangible cultural heritage replicas.
\newblock In \emph{Proceedings of the Eurographics Workshop on Graphics and
  Cultural Heritage} (2014), Eurographics Association, pp.~145--154.

\bibitem[AMG{\etalchar{*}}18]{Alderighi:2018:MCD}
\textsc{Alderighi T., Malomo L., Giorgi D., Pietroni N., Bickel B., Cignoni
  P.}:
\newblock Metamolds: Computational design of silicone molds.
\newblock \emph{ACM Trans. Graph. 37}, 4 (July 2018), 136:1--136:13.
\newblock URL: \url{http://doi.acm.org/10.1145/3197517.3201381}, \href
  {http://dx.doi.org/10.1145/3197517.3201381}
  {\path{doi:10.1145/3197517.3201381}}.

\bibitem[BCMP18]{BCMP18}
\textsc{Bickel B., Cignoni P., Malomo L., Pietroni N.}:
\newblock State of the art on stylized fabrication.
\newblock \emph{Computer Graphics Forum 37} (2018).
\newblock URL: \url{http://vcg.isti.cnr.it/Publications/2018/BCMP18}.

\bibitem[BK04]{boykov:2004:experimental}
\textsc{Boykov Y., Kolmogorov V.}:
\newblock An experimental comparison of min-cut/max-flow algorithms for energy
  minimization in vision.
\newblock \emph{IEEE Trans. Pattern Anal. Mach. Intell. 26}, 9 (2004),
  1124--1137.

\bibitem[BWBSH14]{SpinIt:Baecher:2014}
\textsc{B\"acher M., Whiting E., Bickel B., Sorkine-Hornung O.}:
\newblock {Spin-It}: Optimizing moment of inertia for spinnable objects.
\newblock \emph{ACM Trans. Graph. (Proc. SIGGRAPH 2014) 33}, 4 (2014),
  96:1--96:10.

\bibitem[CLF{\etalchar{*}}18]{Chen:2018:FUB}
\textsc{Chen X., Li H., Fu C.-W., Zhang H., Cohen-Or D., Chen B.}:
\newblock 3d fabrication with universal building blocks and pyramidal shells.
\newblock \emph{ACM Trans. Graph. 37}, 6 (Dec. 2018), 189:1--189:15.
\newblock URL: \url{http://doi.acm.org/10.1145/3272127.3275033}, \href
  {http://dx.doi.org/10.1145/3272127.3275033}
  {\path{doi:10.1145/3272127.3275033}}.

\bibitem[CPMS14]{Cignoni:2014:FMJ}
\textsc{Cignoni P., Pietroni N., Malomo L., Scopigno R.}:
\newblock Field-aligned mesh joinery.
\newblock \emph{ACM Trans. Graph. 33}, 1 (2014), 11:1--11:12.

\bibitem[CV95]{cortes1995support}
\textsc{Cortes C., Vapnik V.}:
\newblock Support-vector networks.
\newblock \emph{Mach. Learn. 20}, 3 (1995), 273--297.

\bibitem[Dav07]{davis2007mathematics}
\textsc{Davis T.}:
\newblock The mathematics of zome, 2007.

\bibitem[HJJ03]{Henderson:2003:SA}
\textsc{Henderson D., Jacobson S.~H., Johnson A.~W.}:
\newblock \emph{The Theory and Practice of Simulated Annealing}.
\newblock Springer US, 2003.

\bibitem[HLZCO14]{Hu:2014:APS}
\textsc{Hu R., Li H., Zhang H., Cohen-Or D.}:
\newblock Approximate pyramidal shape decomposition.
\newblock \emph{ACM Trans. Graph. 33}, 6 (2014), 213:1--213:12.

\bibitem[HZH{\etalchar{*}}16]{Huang:2016:FRF}
\textsc{Huang Y., Zhang J., Hu X., Song G., Liu Z., Yu L., Liu L.}:
\newblock Framefab: Robotic fabrication of frame shapes.
\newblock \emph{ACM Trans. Graph. 35}, 6 (2016), 224:1--224:11.

\bibitem[LBRM12]{Luo:2012:CPM}
\textsc{Luo L., Baran I., Rusinkiewicz S., Matusik W.}:
\newblock Chopper: Partitioning models into {3D}-printable parts.
\newblock \emph{ACM Trans. Graph. (Proc. SIGGRAPH Asia 2012) 31}, 6 (2012).

\bibitem[LSZ{\etalchar{*}}14]{Lu:2014:BSW}
\textsc{Lu L., Sharf A., Zhao H., Wei Y., Fan Q., Chen X., Savoye Y., Tu C.,
  Cohen-Or D., Chen B.}:
\newblock Build-to-last: Strength to weight {3D} printed objects.
\newblock \emph{ACM Trans. Graph. 33}, 4 (2014), 97:1--97:10.

\bibitem[LYH{\etalchar{*}}15]{Luo:2015:LOL}
\textsc{Luo S.-J., Yue Y., Huang C.-K., Chung Y.-H., Imai S., Nishita T., Chen
  B.-Y.}:
\newblock Legolization: Optimizing {LEGO} designs.
\newblock \emph{ACM Trans. Graph. (Proc. SIGGRAPH Asia 2015) 34}, 6 (2015),
  222:1--222:12.

\bibitem[MIG{\etalchar{*}}14]{Mueller:2014:WPP}
\textsc{Mueller S., Im S., Gurevich S., Teibrich A., Pfisterer L.,
  Guimbreti\`{e}re F., Baudisch P.}:
\newblock Wireprint: 3{D} printed previews for fast prototyping.
\newblock In \emph{Proc. ACM UIST '14} (2014), pp.~273--280.

\bibitem[PWLSH13]{Prevost:MIS:2013}
\textsc{Pr\'evost R., Whiting E., Lefebvre S., Sorkine-Hornung O.}:
\newblock {Make It Stand}: Balancing shapes for 3{D} fabrication.
\newblock \emph{ACM Trans. Graph. (Proc. SIGGRAPH 2013) 32}, 4 (2013),
  81:1--81:10.

\bibitem[SBM16]{Shamir:2016:CTP}
\textsc{Shamir A., Bickel B., Matusik W.}:
\newblock Computational tools for {3D} printing.
\newblock In \emph{ACM SIGGRAPH 2016 Courses} (2016), pp.~9:1--9:34.

\bibitem[SDW{\etalchar{*}}16]{Song-2016-CofiFab}
\textsc{Song P., Deng B., Wang Z., Dong Z., Li W., Fu C.-W., Liu L.}:
\newblock {CofiFab}: Coarse-to-fine fabrication of large {3D} objects.
\newblock \emph{ACM Trans. Graph. (Proc. SIGGRAPH 2016) 35}, 4 (2016).

\bibitem[SFCO12]{Song-2012-InterCubes}
\textsc{Song P., Fu C.-W., Cohen-Or D.}:
\newblock Recursive interlocking puzzles.
\newblock \emph{ACM Trans. Graph. (SIGGRAPH Asia 2012)} (2012), 128:1--128:10.

\bibitem[SFS02]{Salamon:2002:SA}
\textsc{Salamon P., Frost R., Sibani P.}:
\newblock \emph{Facts, conjectures, and improvements for simulated annealing}.
\newblock Society for Industrial and Applied Mathematics, 2002.

\bibitem[UPSW16]{Umetani:2016:PIR}
\textsc{Umetani N., Panotopoulou A., Schmidt R., Whiting E.}:
\newblock Printone: Interactive resonance simulation for free-form print-wind
  instrument design.
\newblock \emph{ACM Trans. Graph. 35}, 6 (2016), 184:1--184:14.

\bibitem[VGB{\etalchar{*}}14]{Vanek:2014:PMVO}
\textsc{Vanek J., Galicia J. A.~G., Benes B., Mech R., Carr N.~A., Stava O.,
  Miller G. S.~P.}:
\newblock {PackMerger}: A {3D} print volume optimizer.
\newblock \emph{Comput. Graph. Forum 33}, 6 (2014), 322--332.

\bibitem[WPGM16]{Wu:2016:PAM}
\textsc{Wu R., Peng H., Guimbreti\`{e}re F., Marschner S.}:
\newblock Printing arbitrary meshes with a {5DOF} wireframe printer.
\newblock \emph{ACM Trans. Graph. 35}, 4 (2016), 101:1--101:9.

\bibitem[WWY{\etalchar{*}}13]{Wang:2013:CPO}
\textsc{Wang W., Wang T.~Y., Yang Z., Liu L., Tong X., Tong W., Deng J., Chen
  F., Liu X.}:
\newblock Cost-effective printing of 3{D} objects with skin-frame structures.
\newblock \emph{ACM Trans. Graph. 32}, 6 (2013), 177:1--177:10.

\bibitem[WZK16]{wang2016improved}
\textsc{Wang W.~M., Zanni C., Kobbelt L.}:
\newblock Improved surface quality in {3D} printing by optimizing the printing
  direction.
\newblock \emph{Comput. Graph. Forum (Proc. EG 2016) 35}, 2 (2016), 59--70.

\bibitem[YCL{\etalchar{*}}15]{Yao:2015:LPP}
\textsc{Yao M., Chen Z., Luo L., Wang R., Wang H.}:
\newblock Level-set-based partitioning and packing optimization of a printable
  model.
\newblock \emph{ACM Trans. Graph. 34}, 6 (2015), 214:1--214:11.

\bibitem[ZK14]{zimmer:2014:tvcg}
\textsc{Zimmer H., Kobbelt L.}:
\newblock Zometool rationalization of freeform surfaces.
\newblock \emph{IEEE Trans. Vis. Comput. Graph. 20}, 10 (2014), 1461--1473.

\bibitem[ZLAK14]{zimmer:2014:Zometool}
\textsc{Zimmer H., Lafarge F., Alliez P., Kobbelt L.}:
\newblock Zometool shape approximation.
\newblock \emph{Graph. Models 76}, 5 (2014), 390--401.

\bibitem[ZPZ13]{Zhou:2013:WSA}
\textsc{Zhou Q., Panetta J., Zorin D.}:
\newblock Worst-case structural analysis.
\newblock \emph{ACM Trans. Graph. 32}, 4 (2013), 137:1--137:12.

\end{thebibliography}

\end{document}



\maketitle
\section{More results}
We show more results here.


\begin{table*}
\centering
\resizebox{0.965\linewidth}{!} {
\begin{tabular}{c} 
\includegraphics{figs/MOAI.pdf} \\
\includegraphics{figs/Squirrel.pdf}\\
\includegraphics{figs/doraemon.pdf} \\
\end{tabular}
}
\caption{}
\label{tab:result_ZomeFab_file_1}
\end{table*}

\begin{table*}
\centering
\resizebox{1.\linewidth}{!} {
\begin{tabular}{c} 
\includegraphics{figs/owl.pdf}\\
\includegraphics{figs/pig.pdf} \\
\includegraphics{figs/slime_high.pdf}\\
\includegraphics{figs/lion.pdf}\\
\includegraphics{figs/bunny_150.pdf}\\
\end{tabular}
}
\caption{}
\label{tab:result_ZomeFab_file_2}
\end{table*}
